\documentclass[aps,epsf,amsmath,amssymb,preprint]{revtex4}
\usepackage{epsfig}

\begin{document}


\title{Differences between discontinuous and continuous
soft-core attractive potentials: the appearance of density anomaly}

\author{Giancarlo Franzese}
\affiliation{Departament de Fisica Fonamental, Facultat de Fisica,
Universitat de Barcelona, Diagonal 647, 08028, Barcelona, Spain.}
\email{gfranzese@ub.edu}

\begin{abstract}
Soft-core attractive potentials can give rise to a phase diagram with
three fluid phases at different densities (gas, low-density liquid and
high-density liquid), separated by first order
phase transition lines ending in critical points. Experiments show
a phase diagram with these features for phosphorous and
triphenyl-phosphite. Liquid-liquid phase transition 
could be relevant for water, silica, liquid metals,
colloids and protein solutions, among others. 
Here we compare two potentials
with short-range soft-core repulsion and narrow attraction. One of
them is a squared potential that is known to have liquid-liquid phase
transition, ending in a critical point, and no anomaly in
density. The normal, monotonic, behavior of density for isobaric
cooling is surprising if compared with molecular liquids, such as water,
where a hypothetical critical point is proposed as rationale for the
anomalous behavior of density. The second potential
is a continuous version of the first. We show
that the phase diagram associated to this new potential has, not only
the liquid-liquid phase transition, but also the density anomaly.
Our result, therefore, shows that the behavior in density is strongly
dependent on the derivative of the potential.\\


\end{abstract}

\maketitle

\section{Introduction}

Increasing experimental, theoretical and numerical evidences are
showing that single-component systems can have more than the two
commonly known fluid phases: gas and liquid. There is no
thermodynamic inconsistency, indeed, in hypnotizing the 
possible existence of a phase transition between two (or more) liquids
at different densities. 
Direct experimental evidences of this
phenomenon have been observed in liquid phosphorous
\cite{Katayama,Monaco} and triphenyl phosphite \cite{Kurita}, while 
consistent data exists for water
\cite{mishima_2000,bellisent,soper,SuzukiMishima2002,Loerting01}, silica
\cite{ANGELL,Lacks}, 
aluminate liquids \cite{McMillan}, selenium
\cite{Brazhkin98}, and cobalt \cite{Vasin}, among others
\cite{McMillan2}. Simulations predict a liquid-liquid critical point
for supercooled water
\cite{Poole,fs,fs1,FMS,fs2,slt,Brovchenko}, 
phosphorous
\cite{MORISHITA}, 
supercooled silica \cite{ANGELL,Saika-Voivod,SHRI_PP}, and hydrogen
\cite{hydrogen}.
For other substances, such as carbon, literature is contradictory
\cite{Thiel,G,Ghiringhelli}.

In the last years, several models have been proposed to understand the
origin of the liquid-liquid phase transition. 
Within this context a large interest has been focused on isotropic
core-softened  models \cite{debenerev03}. The reason is
twofold. On the one hand, within acceptable limitations, 
they are models for 
a variety of systems including liquid metals, metallic
mixtures, electrolytes, colloids and protein solution, as well as
anomalous liquids, like water and silica
\cite{hs,Deb1,SLK,SA,L75,DRB91,SHG93,Deb2,ssbs,jagla,wilding,Caballero}.
On the other hand, their simple definition encourages to investigate
the possibility of an intriguing relation between the liquid-liquid
phase transition and the anomalies in specific observables, such as
the density or the diffusivity, in network-forming liquids, like water
and silica. 

Different kinds of isotropic potentials with soft-core have been
proposed. In many of them, with the soft-core given by a Gaussian core
\cite{StillingerJCP78} or a ramp, the
presence of a liquid-liquid phase transition is observed together with
anomalies in the density, in the diffusion and in the structure
\cite{jagla,wilding,stanley,sharma,camp,esposito,deoliveira,mittal}.
This is consistent with what is proposed for water on the basis of
simulations and theoretical calculations for specific models
\cite{Poole,FMS,debenerev03,Brovchenko}.

In other attractive isotropic potentials, with the soft-core given by a square
shoulder, the liquid-liquid 
phase transition occurs in absence of density anomaly
\cite{nature,fmsbs,sbfms,mfsbs}. Nevertheless, an asymptotic tendency
to the density anomaly is observed in three dimensions \cite{fmsbs},
and thermodynamic and dynamic anomalies has been found in related
models in two dimensions on lattices \cite{barbosa}
or off-lattice \cite{buldyrev,camp}.

Here we focus on the study of a potential that is the smooth version
of the one with square shoulder and attractive well which we studied
in Ref.~\cite{nature}.
We find that the new continuous potential has a liquid-liquid phase
transition, as well as the previous squared potential,
consistent with the minimal difference between the two models.
Nevertheless, the difference is enough to cause the
appearance of the density anomaly. Hence, the anomalous behavior is 
strongly dependent on the fine details of the potential, including its  
derivative. 

\section{The continuous potential}

We consider the following isotropic pairwise potential

\begin{equation}
U_0(r)=
\frac{U_R}
{1+\exp\left[\Delta\left( r-R_R \right)/a\right]}
-
U_A \exp\left[ -\frac{(r-R_A)^2}{ 2\delta_A^2}\right] 
+
\left( \frac{a}{r}\right)^{24} ~,
\label{pot}
\end{equation}
where 
$U_R$ and $U_A$ are
the energy of the repulsive shoulder and of the attractive well, respectively;
$a$, $R_R$ and $R_A$ are the hard-core distance, the
repulsive average radius and the distance of the attractive minimum,
respectively; 
$\Delta$ is a parameter related to the slope of the potential at $R_R$;
$\delta_A^2$ is the variance of the Gaussian centered at $R_A$.

\begin{figure}[ht]
\begin{center}
\includegraphics[clip=true,scale=0.6]{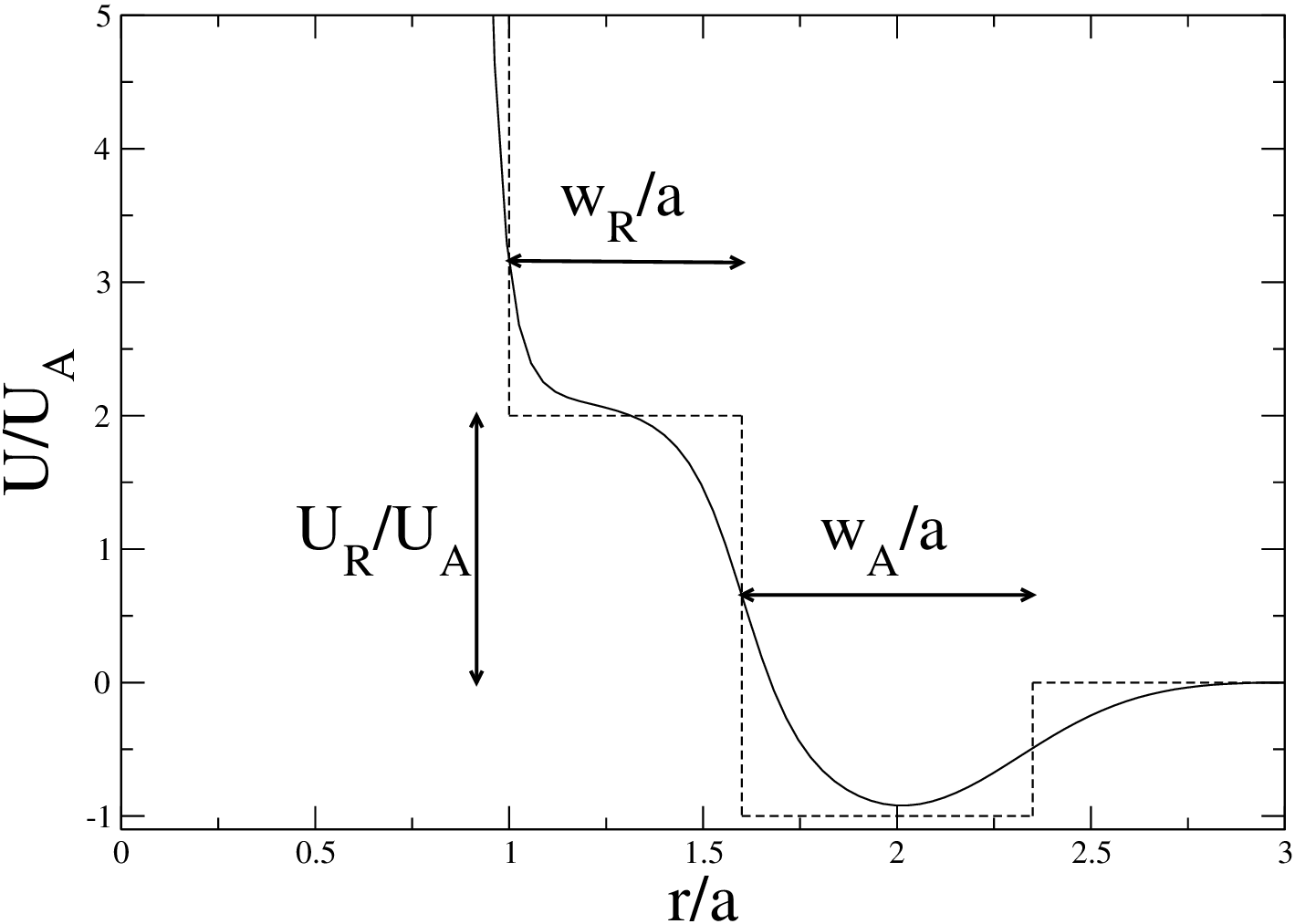}
\end{center}
\caption{Isotropic pairwise interaction potential in Eq.~(\ref{pot}),
for the parameters in the text (continuous line), compared with the
squared potential (dashed line) of Refs.~\cite{nature,fmsbs,sbfms,mfsbs} for 
$U_R/U_A=2$, repulsive width $w_R/a=0.6$ and attractive width
$w_A/a=0.75$. The two potentials have the same $U_R/U_A$, the same 
repulsive range $w_R= (R_R-a)$ and the same 
attractive width $w_A= 2 \delta_A \sqrt{2 \ln(2)}$.
The continuous potential crosses the lines
of the shoulder and well of the squared potential approximately in
the middle.}
\label{potential}
\end{figure}

For the set of parameters 
$U_R/U_A=2$,
$R_R/a=1.6$ 
$R_A/a=2$,
$\Delta=15$,
$(\delta_A/a)^2=0.1$, 
at distance $r/a=3$ the value of the potential is $U_0/U_A\simeq
-6.7\times 10^{-3}$, with $U_0\to 0$ for $r/a>3$. 
Hence, to reduce the computational effort, we set a potential cutoff
at $r/a=3$ and add a constant and a linear term to the potential in 
order to have both the potential and its $r$-derivative equal to zero at
the cutoff.

The resulting potential is 
\begin{equation}
U(r)=U_0(r)+C U_A +B \frac{r}{a}
\end{equation}
where, for our choice of parameters and cutoff distance $r_c/a=3$,
the dimensionless constants are set to
$C=0.208876$ 
and 
$B=-0.0673794$.
The approximate attractive width, calculated as the
width at half height of the Gaussian function 
defining the
attractive well,
is
$w_A= 2 \delta_A \sqrt{2 \ln(2)} 
\simeq 0.75 ~a$.

The continuous potential smoothly follows the corresponding squared one
(Figure~\ref{potential}).
While the continuous potential is defined by five parameters
(plus the cutoff distance and two additional constants),
the squared potential is
defined by only three parameters: $U_R/U_A$, the repulsive width
$w_R/a=(R_R-a)/a$ of the shoulder and the attractive width $w_A/a$ of the
well.
A straightforward correspondence between the two types of potential
can be established by choosing the same $U_R/U_A$, the same $w_R= (R_R-a)$
and $w_A= 2 \delta_A \sqrt{2 \ln(2)}$, respectively.

The squared potential is more appropriate for a systematic
study of the phase diagram changes for different combinations of
parameters. This analysis has been performed in
Refs.~\cite{fmsbs,sbfms,mfsbs}.  
The continuous potential, on the other hand,
resembles effective pair potentials obtained from the second-order
pseudo-potential perturbation theory for liquid alkaline-earth metals,
such as magnesium, calcium, strontium and barium near the melting point
\cite{WaxPRB00,DebenedettiJPC91}, 
or effective potentials for colloidal mono-layers at
the water-air interface given by inversion of structural data
\cite{QuesadaJCP01}. 

Moreover, the angle-average of several asymmetric water potentials
give rise to isotropic soft-core effective interactions, 
such as the one studied in Refs. 
\cite{ChoPRL,NetzPhysA04}.
Hence, the continuous potential
appears to be a more realistic model for
different problems.

\section{Molecular Dynamics Simulations} 

We perform Molecular Dynamics (MD) simulations in the $NVT$ ensemble for 
$N=1372$ particles of unit mass $m$ in 
a cubic box of volume $V$ with periodic boundary
conditions at temperature $T$. We use standard MD technique for
continuous potentials, with velocity Verlet integrator \cite{AT}.
We keep $T$ constant by rescaling the velocities at each time step by
a factor of $(T/{\cal T})^{1/2}$ where $\cal T$ is the current kinetic
temperature and $T$ is the desired thermodynamic temperature \cite{AT}.
We
calculate the pressure $P$ by means of its expression in terms of the
second virial coefficient \cite{AT}. 

We consider 28 densities $\rho\equiv N/V$
in the range $0.05\leq \rho a^3\leq 0.33$ and 5 temperatures in the range
$0.4\leq k_BT/U_A \leq 1.9$, where $k_B$ is the Boltzmann constant. 
For each state point we run 8 independent simulations from different
configurations equilibrated at $T+\delta T$, with $k_B\delta
T/U_A=0.1$ for the state points with $0.4\leq k_BT/U_A \leq 0.8$,
and with $k_B\delta T/U_A=0.2$ for the state points with $0.9\leq
k_BT/U_A \leq 1.9$. 

Each run has $10^6$ MD steps, with a time step
$\delta t=3.2 \times 10^{-3}$ times units, where
time is measured in units of $(a^2m/U_A)^{1/2}$ 
(of the order of $\approx 2.1\times 10^{-12}$s for for argon-like atoms 
and $\approx 1.7\times 10^{-12}$s for water-like molecules).
We record pressure and potential energy every $\Delta t=100$ MD 
steps. For the calculation of the averages we consider only the data
corresponding to states that are stable within the simulation time, as
follows from the analysis of the time series of pressure, average
temperature, and potential energy. Usually an equilibration time of 
$10^5$ MD steps is enough to reach a stable state.
We average the data binned in $10^3$ consecutive MD steps, a time that
we test to be enough to 
decorrelate the data for the considered state points.

For densities $0.275\leq \rho a^3\leq 0.330$ at $k_BT/U_A=0.5$, and 
for $\rho a^3\geq 0.21$ at $k_BT/U_A=0.4$, we
observe spontaneous crystallization
during the simulation time. 
To get data for the metastable liquid phase in these cases, instead of
annealing from an equilibrium configuration at $T+\delta T$, 
we perform instantaneous quenches from configurations
equilibrated at $k_BT/U_A=0.8$. Next, we analyze the evolution in time
of the structure factor, as described in details in Ref.~\cite{fmsbs}, and
we use for the averages only the data corresponding to the metastable
liquid state. We accumulate statistics for these state points up to
reach a number of independent metastable configurations comparable to
the number used for the other state points.
We try to analyze in the same way also data for
quenches to $k_BT/U_A=0.3$ and lower $T$, but the
crystal nucleation time observed in these cases is too small to allow
us reliable estimations.

\section{The phase diagram}

\begin{figure}[ht]
\begin{center}
\includegraphics[clip=true,scale=0.6]{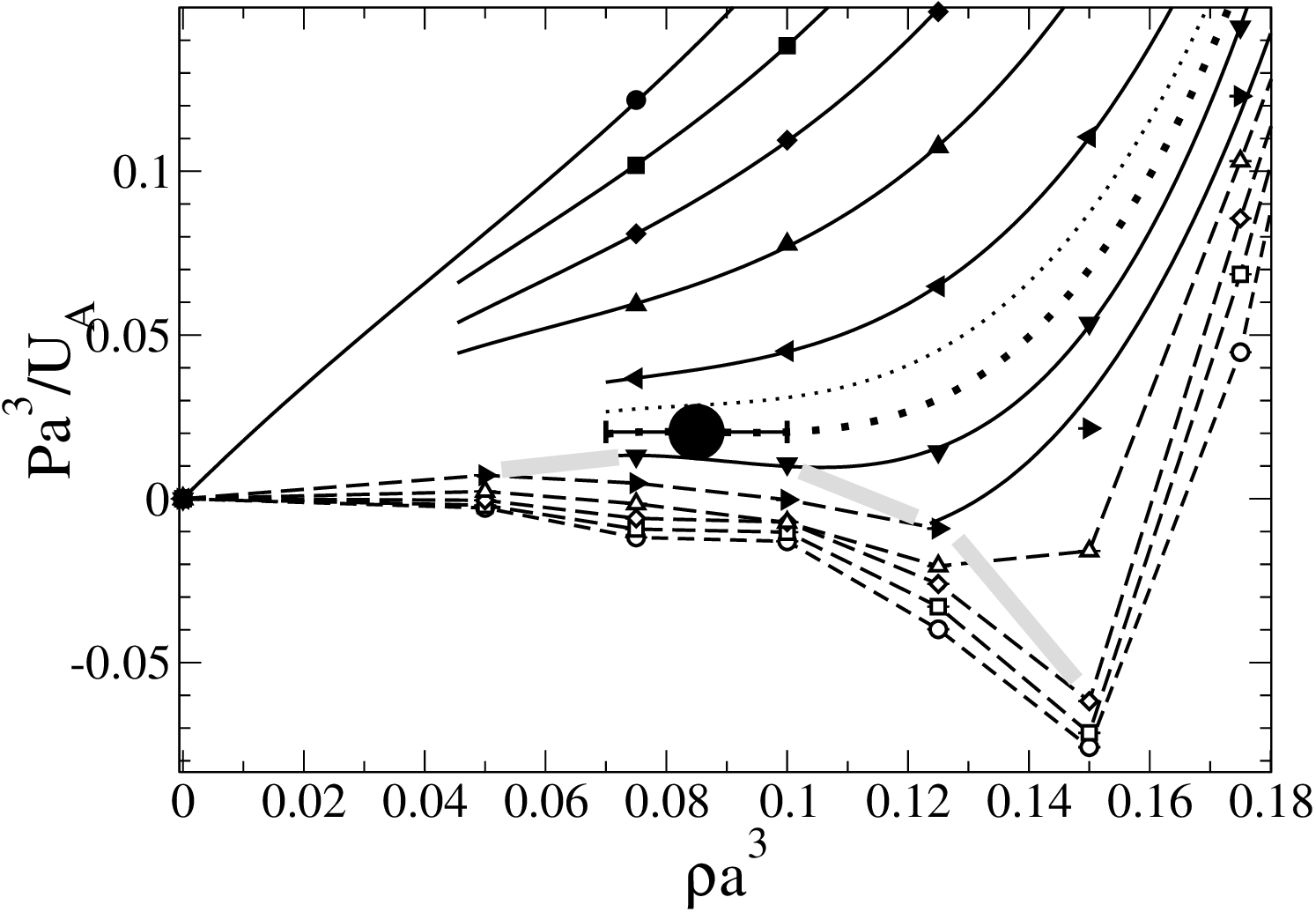}
\end{center}
\caption{Low density phase diagram with the estimate of the
gas--liquid critical point ${C_1}$ (large full circle) at
$\rho_{C_1} a^3=0.08\pm 0.02$,
$P_{C_1}a^3/U_A=0.021\pm 0.008$, and
$k_BT_{C_1}/U_A=0.96\pm 0.06$. 
Symbols are state points calculated with MD in the pressure-density
($P$--$\rho$) plane for, from top to bottom, $k_BT/U_A=1.9$, 1.7, 1.5,
1.3, 1.1, 0.9, 0.8, 0.7, 0.6, 0.5, 0.4. 
Full lines are cubic interpolations of the isotherms. 
Dashed lines are a guide for the eyes connecting symbols for the
isotherms at which the cubic interpolation is not reliable. 
Dotted lines are liner interpolations between the isotherm at
$k_BT/U_A=1.1$ and the isotherm at $k_BT/U_A=0.9$ for (upper thin
dotted line) $k_BT/U_A=1.02$ and (lower thick dotted line)
$k_BT/U_A=0.96$, approximating the critical isotherm.
The thick gray dashed line at $T<T_{C_1}$ is a guide for the eyes
connecting the points of minimum $P$ along the isotherms for
$\rho>\rho_{C_1}$ (i. e. the limit of stability of the liquid with respect
to the gas), and the points of maximum $P$ for $\rho<\rho_{C_1}$ (i. e. the
limit of stability of the gas with respect to the liquid), and
representing an estimate of the spinodal line.
Errors on state points are smaller then symbols size.}
\label{low-rho}
\end{figure}

We analyze the isotherms in the $P$--$\rho$ phase diagram. At low
density (Figure.~\ref{low-rho}) we observe non-monotonic isotherms,
showing a maximum at low $\rho$ and a minimum at higher $\rho$ (the
so-called {\it loop}), typical of a first-order phase transition
between two coexisting phases. 
By joining these minima and maxima we get an
estimate of the two branches of the spinodal line, i. e.  the limit of
stability of one of 
the two coexisting phases with respect to the other. The point where
the two branches of the spinodal line meet (the {\it vertex} of the
spinodal) is, by definition, the
gas-liquid critical point $C_1$, with critical pressure $P_{C_1}$, density
$\rho_{C_1}$ and temperature $T_{C_1}$.
Since, by definition, at $\rho_{C_1}$ the $\rho$-derivative of the
critical isotherm $P(\rho,T_{C_1})$ is zero, we interpolate our data to
find the isotherm that appears to be flat around the approximate
vertex of the spinodal line. Our final estimate of the parameters for
the gas-liquid critical point is 
$\rho_{C_1} a^3=0.08\pm 0.02$,
$P_{C_1}a^3/U_A=0.021\pm 0.008$, and
$k_BT_{C_1}/U_A=0.96\pm 0.06$. 

\begin{figure}[ht]
\begin{center}
\includegraphics[clip=true,scale=0.6]{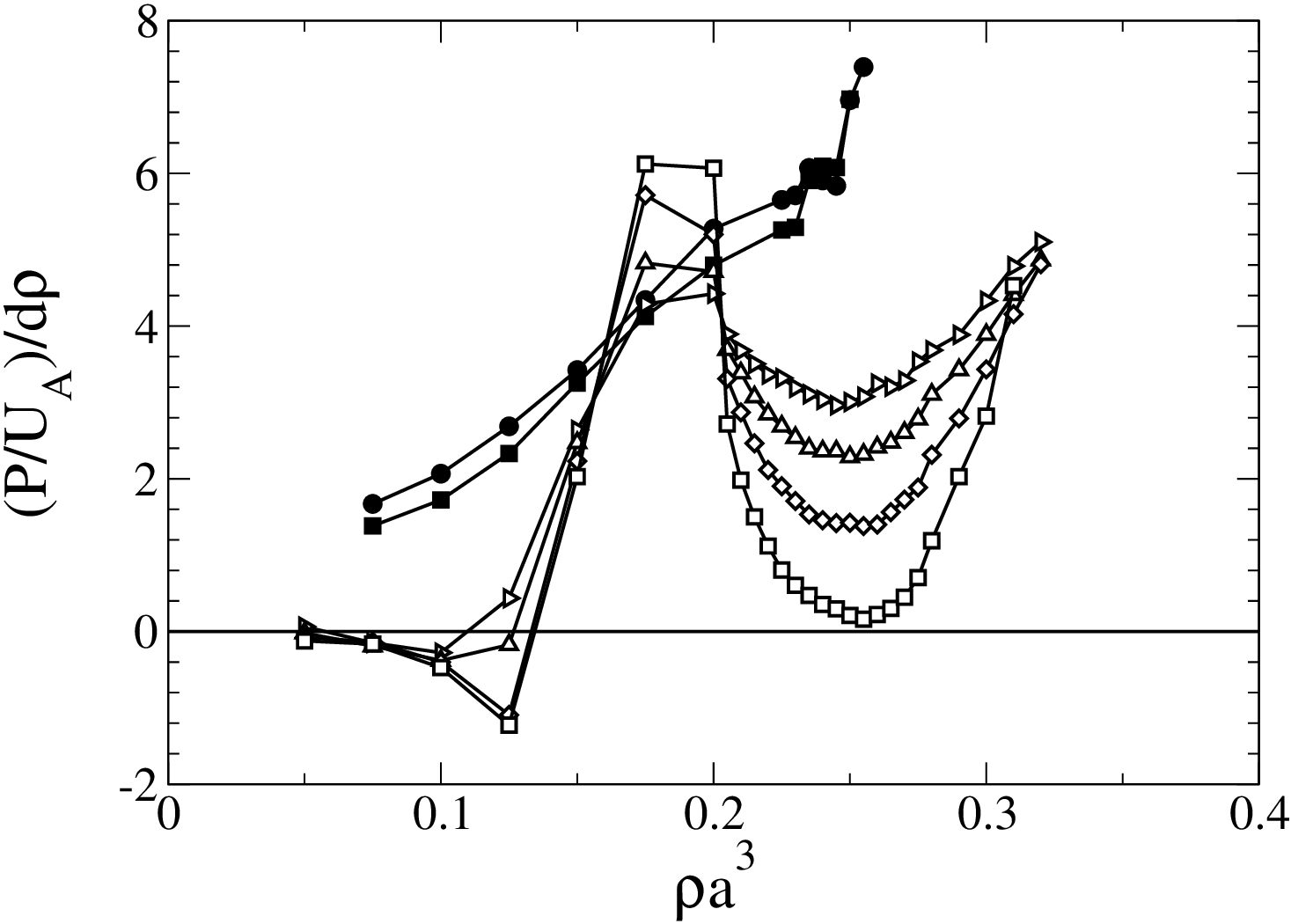}
\end{center}
\caption{Numerical derivative of $P$ with respect to $\rho$ along the
isotherms at, from top to bottom,  $k_BT/U_A=1.9$, 1.7, 0.8,
0.7, 0.6, 0.5. Symbols are as in Figure~\ref{low-rho}.
At high $T$ the derivative increases with $\rho$ as for normal
liquids, while at low $T$ the derivative is non monotonic with a clear
minimum that approaches zero for $k_BT/U_A=0.5$ at density 
$\rho a^3\approx 0.25$.
Errors on state points are smaller then symbols size.
}
\label{rho-derivative}
\end{figure}

\begin{figure}[ht]
\begin{center}
\includegraphics[clip=true,scale=0.32]{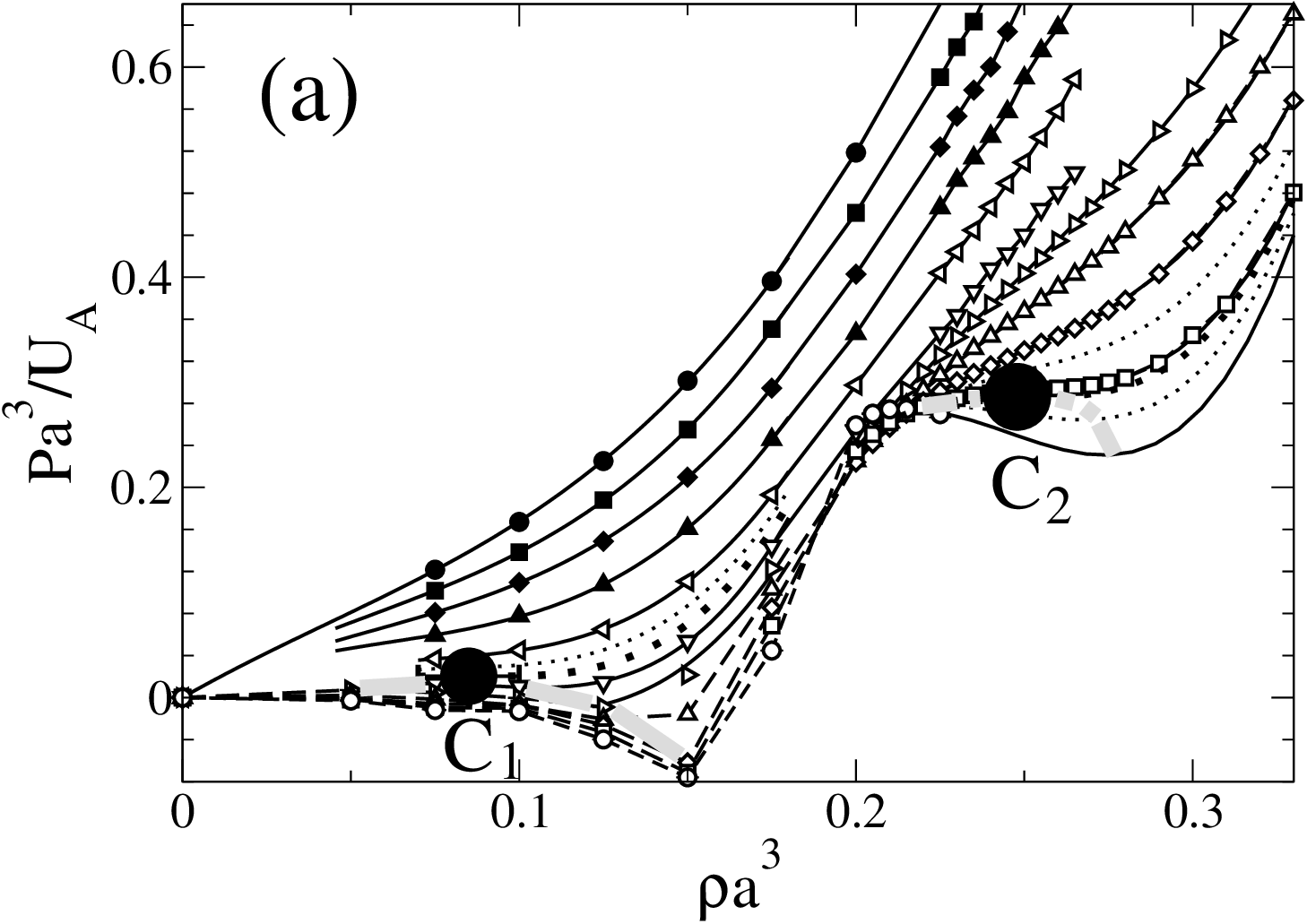}
\includegraphics[clip=true,scale=0.32]{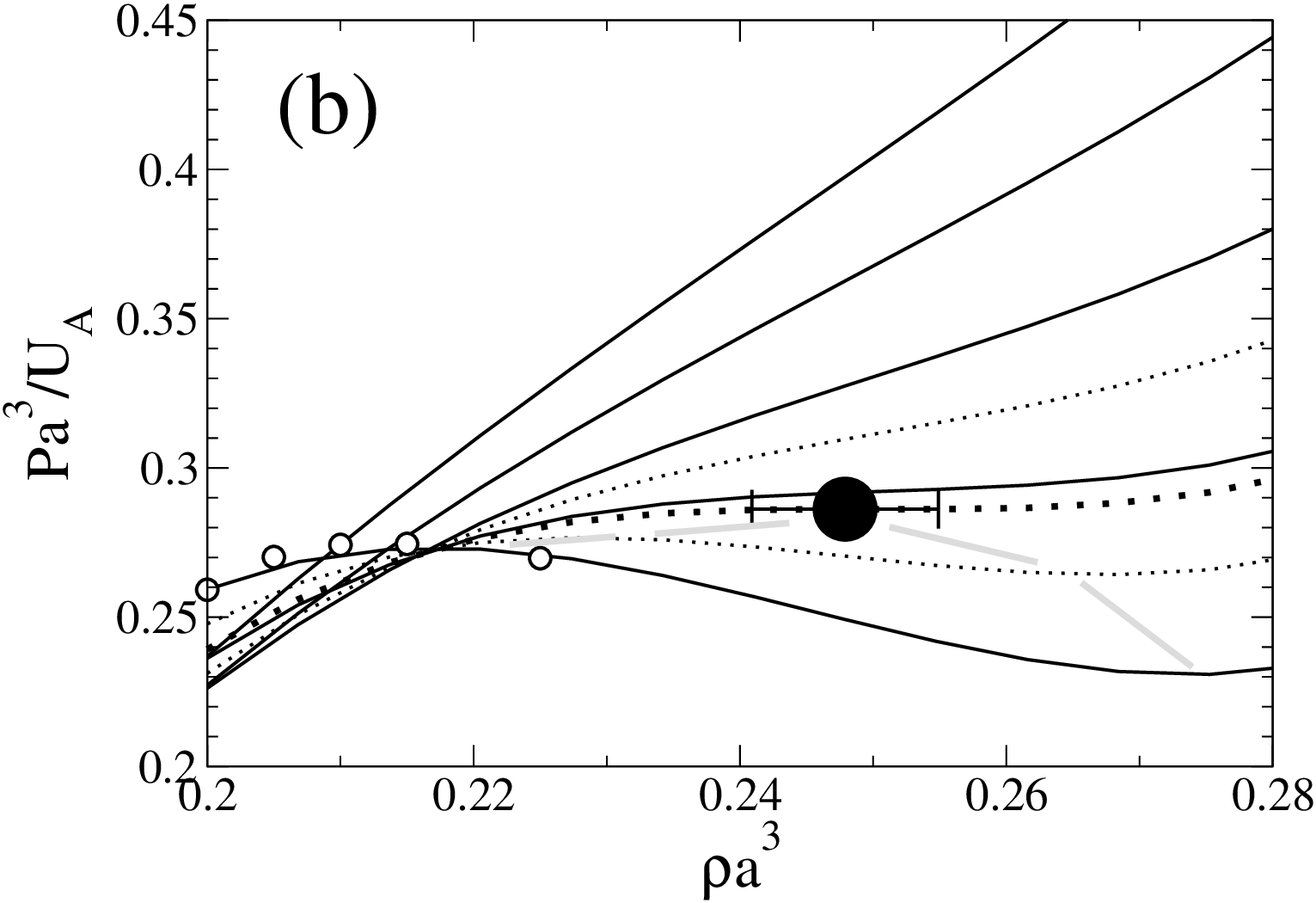}
\end{center}
\caption{(a) Complete phase diagram with the
gas--liquid critical point $C$ (large full circle) as shown in
Figure~\ref{low-rho}, and the liquid--liquid critical point ${C_2}$ 
(large open circle) estimated at
$\rho_{C_2} a^3=0.248\pm 0.007$,
$P_{C_2}a^3/U_A=0.286\pm 0.006$, and
$k_BT_{C_2}/U_A=0.49\pm 0.01$. 
Symbols and lines are as in Figure~\ref{low-rho}.
The lowest solid line below ${C_2}$ is an extrapolation of the data at
$k_BT/U_A=0.4$ in the region of inevitable crystallization, on the
basis of the interpolation of the isotherms at 
higher temperatures. This extrapolation shows very good agreement with
the state points that can be directly calculated at $k_BT/U_A=0.4$
(small open circles).  
In the high-$\rho$ region, from top to bottom, the upper thin dotted line
is the interpolation at $k_BT/U_A=0.55$
between consecutive isotherms, the thick dotted line is the interpolated
critical isotherm, and the lowest thin dotted line is the
interpolations for $k_BT/U_A=0.45$.
The thick gray dashed line at $T<T_{C_2}$ is a guide for the eyes
connecting the points of minima and maxima of $P$ along the isotherms 
representing an estimate of the liquid-liquid spinodal line.
(b) The high-$\rho$ portion of the phase diagram, around the
liquid-liquid critical point $C_2$. Symbols and lines are as in the
previous panel. For sake of clarity we show only the lines
corresponding at the interpolations of the data along the isotherms for  
$0.4 \leq k_BT/U_A\leq 0.8$.
The agreement between the data at $k_BT/U_A=0.4$ (open circles) and
the extrapolation of the corresponding isotherm (lowest solid line) on
the basis of the data at higher $T$ is remarkable.
In both panels, when not shown, errors are smaller then symbols size.}
\label{all-rho}
\end{figure}

At higher density we observe that isotherms for $k_BT/U_A\leq 0.8$
show a clear minimum in the derivative of $P$ with respect to $\rho$
(Figure~\ref{rho-derivative}) approaching zero at 
 $k_BT/U_A= 0.5$ and $\rho a^3\approx 0.25$.
If we assume that the isotherms below this temperature behave in a
regular way and with continuity with respect to the isotherms at higher
$T$, then Figure~\ref{rho-derivative}
shows that the isotherm at  $k_BT/U_A= 0.5$  is just above the
critical isotherm, where 
$\left(\partial P/\partial \rho \right)_{T}=0$. 
This is confirmed by the analysis of the state
points equilibrated at $k_BT/U_A= 0.4$ (small open circles in
Figure~\ref{all-rho}) showing a clear maximum in pressure at $\rho
a^3\approx 0.21$, as expected for an isotherm entering into the
instability region delimited by a spinodal line.
Direct observation of the {\it loop} along the isotherm at $k_BTU/U_A=
0.4$ is hindered by the fast nucleation of the crystal, as described
in the previous section. 

To estimate the parameters of this high-density critical point ${C_2}$,
we extrapolate from the available isotherms at higher $T$, the
high-$\rho$ state points for $k_BT/U_A= 0.4$. The resulting curve 
[the lowest continuous line in Figure~\ref{all-rho}(a) and
Figure~\ref{all-rho}(b)]
is in very good agreement with the state points calculated at
$0.20\leq \rho a^3\leq 0.22$, confirming the validity of the
assumption of continuity and regularity with respect to the isotherms
at higher $T$.

Therefore, our simulation show a critical point $C_2$ between two
liquids at high $\rho$ and below the gas--liquid critical
temperature, as seen for the squared soft-core shouldered potential 
studied in Refs.~\cite{nature,fmsbs,sbfms,mfsbs} (Figure~\ref{potential}). 
The high density liquid is metastable with respect to the
spontaneous crystallization and the crystal nucleation process is very
fast around ${C_2}$. 
The interpolation of the data allows us to estimate the critical
parameters of the liquid-liquid critical point ${C_2}$ as 
$\rho_{C_2} a^3=0.248\pm 0.007$,
$P_{C_2}a^3/U_A=0.286\pm 0.006$, and
$k_BT_{C_2}/U_A=0.49\pm 0.01$. 

\section{Comparison with the phase diagram of the squared potential}

We compare the phase diagram found here
with the phase diagram for the corresponding squared potential
(Figure~\ref{potential}) with parameters $U_R/U_A=2$,
$w_R/a=0.6$ and $w_A/a=0.75$. 
The phase diagram of the squared potential with the straightforward
correspondence to the one studied here is not
available in literature, however, in Ref.~\cite{sbfms} we studied a
squared potential with set of parameters $U_R/U_A=2$,
$w_R/a=0.6$ and $w_A/a=0.7$ (see Figure~3 of Ref.~\cite{sbfms}).
Comparison with this potential is shown in Table~\ref{compare}.

\begin{table}[hp]
\caption{\label{compare} Temperatures $T_{C_1}$ and $T_{C_2}$, pressures
$P_{C_1}$ and $P_{C_2}$, and densities $\rho_{C_1}$ and $\rho_{C_2}$,
for the critical points $C_1$ and $C_2$, respectively, computed 
for the continuous potential presented in this work  and the squared
potential with parameters $U_R/U_A=2$, $w_R/a=0.6$ and $w_A/a=0.7$
in Ref.~\cite{sbfms}.}
\begin{ruledtabular}
\begin{tabular}{lrrrrrr}
Potential & $k_BT_{C_1}/U_A$ & $a^3P_{C_1}/U_A$ &
    $a^3\rho_{C_1}$
    & $k_BT_{C_2}/U_A$ & $a^3P_{C_2}/U_A$ &
    $a^3\rho_{C_2}$ \\
\hline
Continuous   &$0.96\pm 0.06$ &$0.021\pm 0.008$ &$0.08\pm 0.02$
        &$0.49\pm 0.01$ &$0.286\pm 0.006$  &$0.248\pm 0.007$\\
Squared  &$1.24\pm0.01$ &$0.03\pm0.01$  &$0.09\pm0.02$
        &$0.69\pm0.02$ &$0.11\pm0.02$  &$0.28\pm0.02$\\
\end{tabular}
\end{ruledtabular}
\end{table}

The comparison shows that the  critical
parameters of the gas--liquid critical point $C_1$ are almost the same
in the two cases with lower values for the continuous potential. 
The percentage of variation is 11\% for $\rho_{C_1}$, 
23\%  $T_{C_1}$, and 30\% for $P_{C_1}$.

For the liquid--liquid critical point $C_2$ the parameter $\rho_{C_2}$
of the continuous potential decreases of the same 
percentage 11\% of the critical density of $C_1$, 
and $T_{C_2}$ decreases of 29\%, a percentage comparable to the
decrease of $T_{C_1}$. The situation is different, instead,
for the variation of $P_{C_2}$, that increases of 61\% in the
continuous case with respect to the case of the squared potential.

Therefore, the largest difference between the squared and the
continuous case is on the pressure of the critical points, with a
considerable increase of $P_{C_2}$. These effects are consistent with the
analysis performed in Refs.~\cite{sbfms,mfsbs} for 
a decrease of the attractiveness of a squared potential, as a
consequence of the decrease of the range of
the attractive well $w_A$, or the increase of the repulsive energy
$U_R/U_A$. Panels (a), (b),
(c), (g), (h), (i) of Figure~9 of Ref.~\cite{sbfms}, indeed, show that
a decrease of $w_A$, or an increase of $U_R/U_A$, determines 
the increase of $P_{C_2}$ and the decrease of $T_{C_2}$, 
while the other critical parameters for $C_1$
and $C_2$ are almost unaffected. 

Hence, 
the continuous potential studied here behaves like a squared potential
less attractive than the one that would be in straightforward
correspondence with it. This is consistent, with the observation that
the volume integral of the attractive part of the present potential is
smaller than the volume integral for the corresponding squared
potential, while the volume integral for the repulsive part is
approximately the same in both potentials. 

As predicted in Ref.~\cite{mfsbs}, this continuous potential with two
critical points satisfy the empirical relation 
\begin{equation}
-2\lesssim \frac{1}{U_AV_{SC}}\int_a^\infty U(r) ~ d\vec{r} \lesssim 1 ,
\label{integral}
\end{equation}
where $V_{SC} \equiv \frac{2\pi}{3}R_R^3$.
With the present choice of parameters we find that the integral in
Eq.(\ref{integral}) has a numerical value of $\approx -0.88$,
supporting the idea that the Eq.(\ref{integral}) is a good empirical
relation to predict if an isotropic potential could display a phase diagram
with two critical points.

\section{Density anomaly}

A closer look at the high-density part of the $P$--$\rho$ phase diagram
[Figure~\ref{all-rho}(b)] shows that at low $T$, close to $C_2$, the
isotherms cross at various densities. This is a clear signal of a
density anomaly.

\begin{figure}[ht]
\begin{center}
\includegraphics[clip=true,scale=0.6]{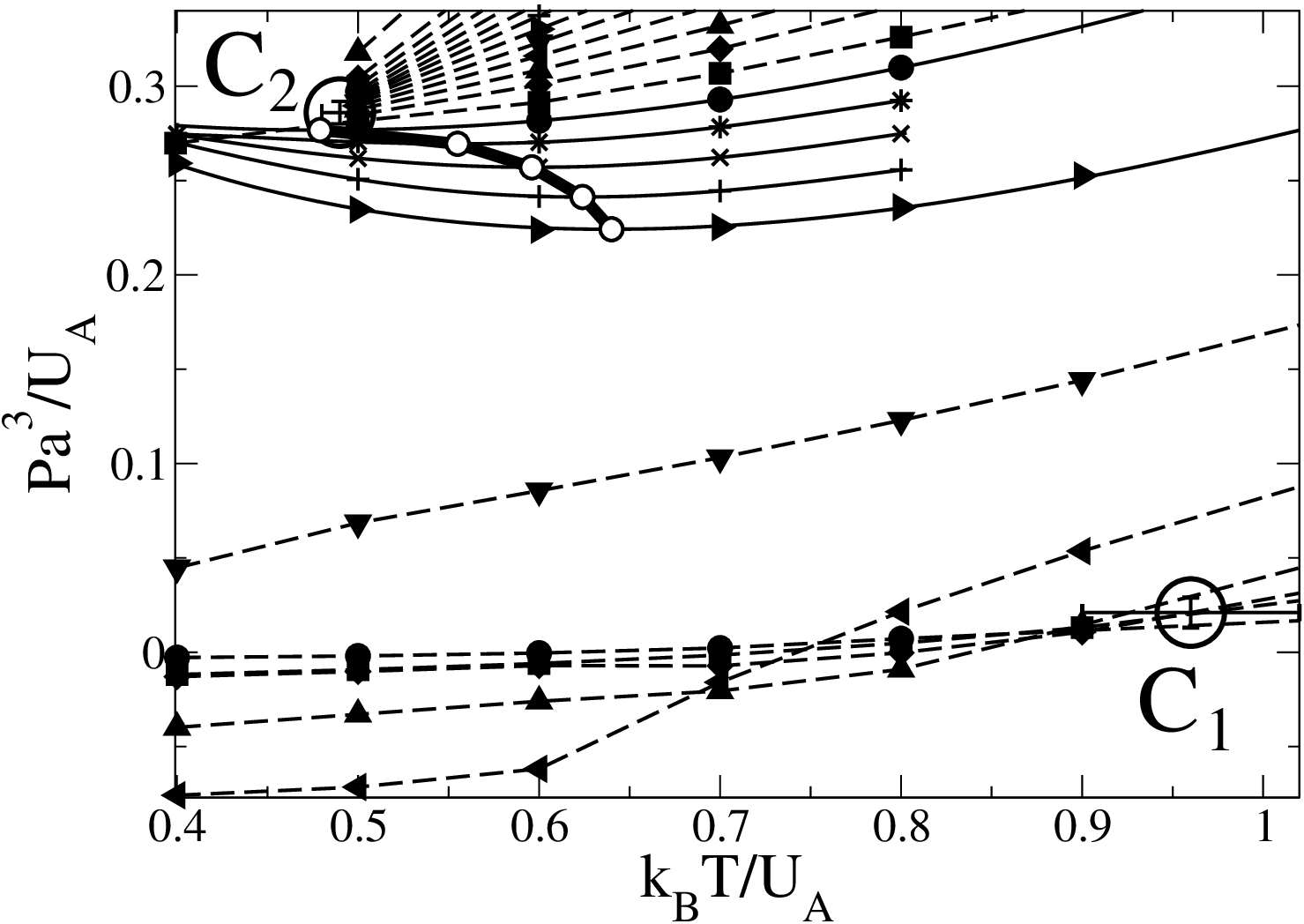}
\end{center}
\caption{The $P$--$T$ phase diagram shows isochores with a clear minimum.
The gas--liquid critical point $C_1$ and the liquid--liquid critical
point $C_2$ are shown on the basis of the estimates in
Figure~\ref{all-rho}.
They are consistent with the crossing occurring at the
highest $T$ along each isochore.
Symbols correspond to isochores for, from bottom to top at
$k_BT/U_A=1$, 
$\rho a^3 = 0.005$, 
0.075, 
0.100, 
0.125, 
0.150, 
0.175, 
0.200, 
0.205, 
0.210, 
0.215, 
0.220, 
0.225, 
0.230, 
0.235,
0.240,
0.245,
0.250,
0.255,
0.260,
0.265,
0.280,
0.290,
0.300,
0.310.
Dashed lines are a guide for the eyes.
Solid lines are a quartic polynomial fit 
between the isochores showing a minimum.
The open circles represent the point where 
$\left(\partial P/\partial T \right)_{\rho}=0$ for the non-monotonic 
isochores. The thick gray line is a guide for the eyes representing an
estimate of the $T_{\rm MD}$ temperature of maximum density at
constant $P$. When not shown, errors are smaller then symbols size.}
\label{P-T}
\end{figure}

Indeed, the crossing of isochores implies 
$\left(\partial P/\partial T \right)_{V}=0$, i.e., 
for one of the Maxwell relations,  
$\left(\partial S/\partial V \right)_{T}=0$, where $S$ is the entropy.
Hence, is
$\left(\partial S/\partial P \right)_{T}=
\left(\partial S/\partial V \right)_{T} 
\left(\partial V/\partial P \right)_{T}=0$,
where we use the fact that 
$\left(\partial V/\partial P \right)_{T}<0$
is finite for liquids.
Hence, for one of the Maxwell relations,  is
$\left(\partial V/\partial T \right)_{P}=0$, 
or equivalently 
$\left(\partial \rho/\partial T \right)_{P}=0$.
Since at high enough $T$ the isobaric density increases on decreasing
$T$, this relation implies that there is a
temperature of maximum density $T_{\rm MD}$ at constant $P$
(Figure~\ref{P-T}).  

\begin{figure}[ht]
\begin{center}
\includegraphics[clip=true,scale=0.6]{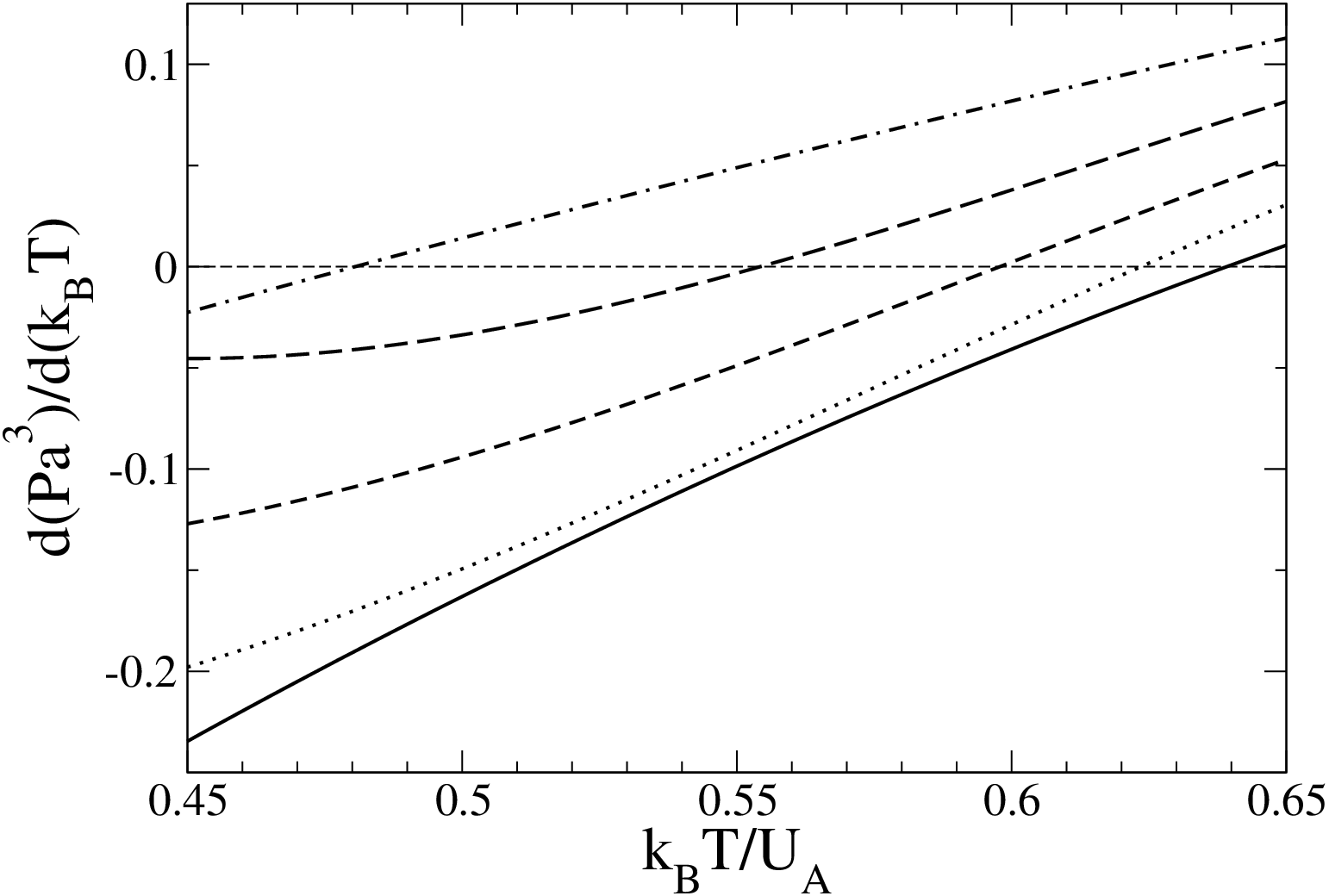}
\end{center}
\caption{Derivative of pressure with respect to temperature along the
isochores at (from bottom to top) $\rho=0.200$, 0.205, 0.210, 0.215,
0.220. Lines are plots of the cubic functions resulting from the
derivative of the best-fit polynomials in Figure~\ref{P-T}, with a
relative errors on the parameters smaller than $10^{-3}$.  The
temperature $T^*$ 
where the derivative is zero is the minimum along the isochore,
with $T^*=T_{\rm MD}$ at the pressure corresponding to the state point 
at the given density and $T^*$.}
\label{dPdT}
\end{figure}

For $T>T_{\rm MD}$ the isobaric density decreases for increasing $T$, as
for normal liquids,
while for  $T<T_{\rm MD}$ it decreases for decreasing $T$, giving
rise to an anomalous behavior in density.
This can be seen in a clear way by studying directly the 
$\left(\partial P/\partial T \right)_{\rho}$ (Figure~\ref{dPdT}).
Comparison with Figure~\ref{all-rho}(b) shows that the isochores with a
minimum, i.e. $\left(\partial P/\partial T \right)_{\rho}=0$,
correspond to the densities where the isotherms 
cross in the $P$--$\rho$ phase diagram.

\section{Discussion and conclusions}

Our results show that the continuous interaction potential presented
here has a maximum in density along the $T_{\rm MD}$ line. This
anomaly in density is typical of network-forming liquids, such as
water or SiO$_2$ \cite{debenerev03,debene-stanley-review,fs2},
and has been observed in other soft-core potentials 
\cite{jagla,wilding,Caballero,stanley,sharma,camp,esposito,deoliveira,mittal}.
In the present model, as well as in those 
for water and other anomalous liquids
\cite{debenerev03}
the $T_{\rm MD}$ line converges toward a liquid--liquid critical point,
i.e. $C_2$ in this work. This observation has been often considered as
an evidence for the fact that the liquid-liquid critical point implies
the presence of a $T_{\rm MD}$ line.

On the other hand, we know that for the squared soft-core potential
the liquid--liquid critical point occurs in absence of the density
anomaly, as shown by the extensive analysis performed in
Ref.~\cite{fmsbs}. As it was observed (see Figure 21 in in
Ref.~\cite{fmsbs}), the squared potential does not satisfy the 
condition for the density anomaly, Eq.(21) in Ref.~\cite{fmsbs},
but approaches that condition in an asymptotic way. Hence, is not clear
if the absence of the $T_{\rm MD}$ is a consequence of the
specific discontinuous shape of the squared potential.

\begin{figure}[ht]
\begin{center}
\includegraphics[clip=true,scale=0.6]{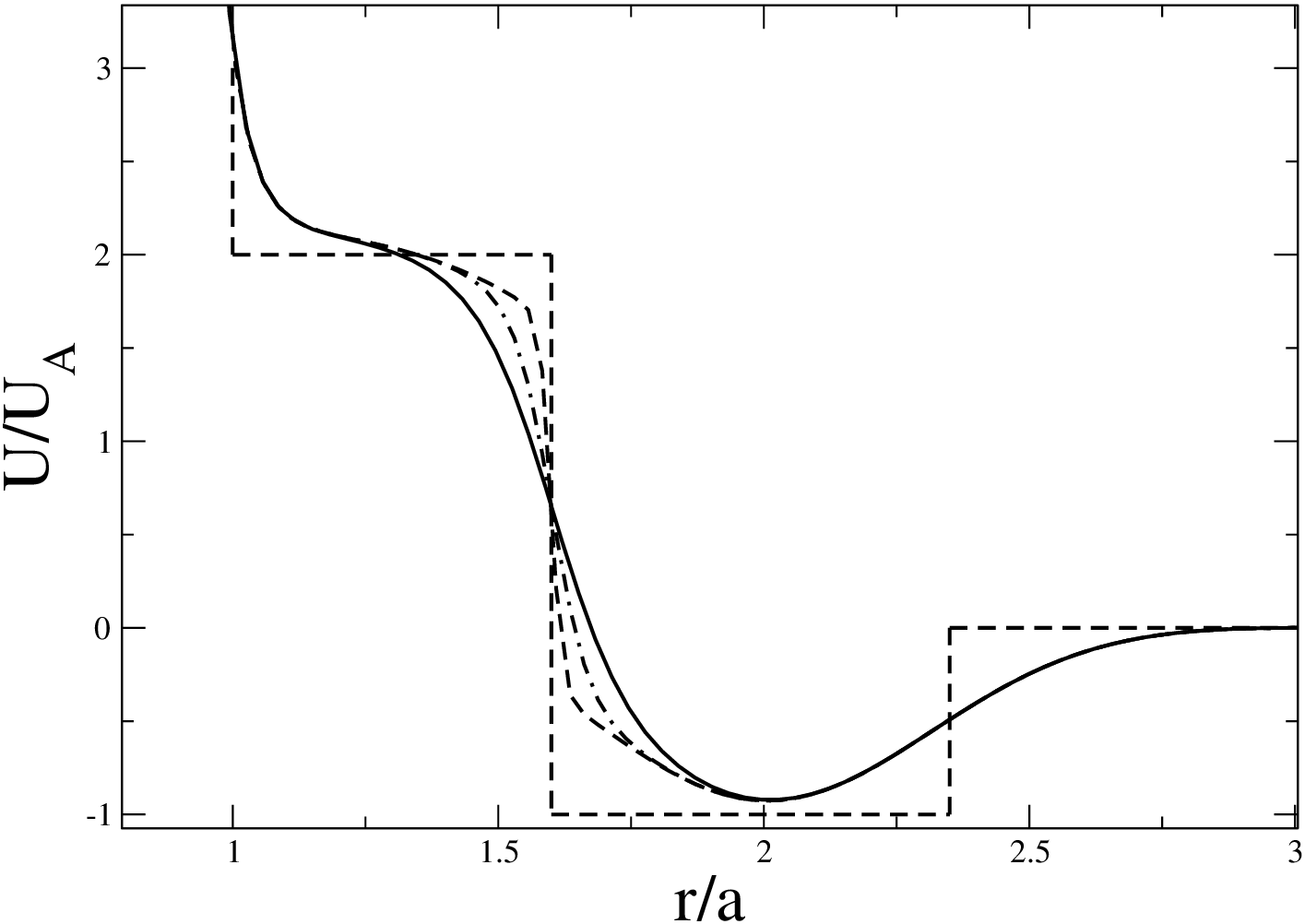}
\end{center}
\caption{The continuous potential can approximate the discontinuity at
the repulsive distance $R_R/a=1.6$ of the squared potential (thin
dashed line) for increasing values of the parameter $\Delta$,
determining the slope at $R_R$.
Continuous line is for $\Delta=15$, studied in this work.
Dash-dotted line is for $\Delta=30$.
Thick dashed line is for $\Delta=100$.}
\label{approx}
\end{figure}

Since the continuous potential presented here can be
tuned to approximate in a close way the discontinuity at $R_R$
for the squared potential (Figure~\ref{approx}), it could be
interesting to study the effect on $T_{\rm MD}$ of the variation
of the parameters $\Delta$ (determining the slope at the $R_R$) of the
continuous potential. Such a study could shed light on the relevant
question about the relation between the liquid-liquid critical point
and the presence of the density anomaly. It is worth to observe here
that the implication of the presence of a liquid-liquid critical point
for network-forming molecular liquids with anomalous density
has been supported by extensive studies on water-like models (see for
example Refs.~\cite{fs,fs1,FMS}).

In conclusion, we have presented here a continuous soft-core potential
with a shape which resembles that of effective interaction
potentials for liquid metals \cite{WaxPRB00} or colloid solutions
\cite{QuesadaJCP01}. 
The model displays a phase diagram with a high-$T$ gas--liquid
critical point $C_1$ at low density, and a low-$T$ liquid--liquid critical
point $C_2$ at high density, occurring at the limit of stability,
within the simulation time, of the
liquid phase with respect to the crystal.

Comparison with the case of the soft-core squared
potentials \cite{nature,fmsbs,sbfms,mfsbs} shows that the continuous
potential is less attractive than its straightforward corresponding
squared potential. The main qualitative difference with the case of
squared potentials is that the continuous potential shows the density
anomaly, typical of network-forming liquids such as water, and
observed in other soft-core potentials
\cite{jagla,wilding,Caballero,stanley,sharma,camp,esposito,deoliveira,mittal}. 
Further investigation is necessary to understand how the appearance of
this anomaly is related to the parameters of the potential.

\subsection*{Acknowledgments}

We thank S. Mossa and F. W. de Sousa Lima that helped in
different stages of this work. 
We acknowledge the allocation of computer resources from INFM
Progetto Calcolo Parallelo at CINECA, Italy.
We thank the Spanish
Ministerio de Educaci\'on y Ciencia for financial support within the 
Programa Ram\'on y Cajal and Grant No. FIS2004-03454.


\begin{thebibliography}{99}

\bibitem{Katayama}
Y. Katayama, T. Mizutani, W. Utsumi, O. Shimomura, M. Ya-makata,
and K. Funakoshi, Nature {\bf 403}, (2000) 170;
Y. Katayama, Y. Inamura, T. Mizutani, M. Yamakata, W. Utsumi, and O. Shimomura,
Science {\bf 306}, (2004) 848.

\bibitem{Monaco}
G. Monaco, S. Falconi, W. A. Crichton, and M. Mezouar, Phys. Rev.
Lett. {\bf 90}, (2003) 255701.

\bibitem{Kurita}
H. Tanaka, R. Kurita, and H. Mataki,
Phys. Rev. Lett. {\bf 92}, (2004) 025701;
R. Kurita and H. Tanaka, Science {\bf 306}, (2004) 845. 

\bibitem{mishima_2000}
O. Mishima and H. E. Stanley, Nature {\bf 396}, (1998) 329;
O. Mishima, Phys. Rev. Lett. {\bf 85}, (2000) 334.

\bibitem{bellisent}
M. C. Bellissent-Funel, Nuovo Cimento {\bf 20D}, 2107 (1998).

\bibitem{soper}
A. K. Soper and M. A. Ricci, Phys. Rev. Lett. {\bf 84}, 2881 (2000).

\bibitem{SuzukiMishima2002}
Y. Suzuki and O. Mishima, 
Nature {\bf 491}, 599 
(2002).

\bibitem{Loerting01}
T. Loerting, C. Salzmann, I. Kohl, E. Mayer, and A. Hallbrucker,
Phys. Chem. Chem. Phys. {\bf 3}, 5355 
(2001).

\bibitem{ANGELL} C. A. Angell, S. Borick, and M. Grabow,
J. Non-Cryst. Solids {\bf 207}, 463 (1996); 
P. H. Poole, M.
Hemmati, and C. A. Angell, Phys. Rev. Lett. {\bf 79}, 2281 (1997).

\bibitem{Lacks}
D. J. Lacks, Phys. Rev. Lett. {\bf 84}, 4629 (2000).

\bibitem{McMillan}
S. Aasland and P. F. McMillan,
Nature {\bf 369}, 633 (1994);
M. C.  Wilding and P. F. McMillan,
J. Non-Cyst. Solids {\bf 293}, 357 (2001).

\bibitem{Brazhkin98}
V. V. Brazhkin, E. L. Gromnitskaya, O. V. Stalgorova, and A. G. Lyapin,
Rev. High Pressure Sci. Technol. {\bf 7}, 1129 (1998).

\bibitem{Vasin}
M. G. Vasin and V. I. Lad\'yanov, Phys. Rev. E {\bf 68}, 051202
(2003).

\bibitem{McMillan2}
P. F. McMillan,
J. Mat. Chem. {\bf 14}, 1506 (2004).

\bibitem{Poole}
P. H. Poole, F. Sciortino, U. Essmann, and H. E. Stanley, Nature {\bf
360}, 324 (1992).

\bibitem{fs} G. Franzese and H. E. Stanley, 
J. Phys.-Cond. Mat. {\bf 14}, 2201 (2002).

\bibitem{fs1} G. Franzese and H. E. Stanley, 
Physica A, {\bf 314}, 
508 
(2002).

\bibitem{FMS}
G. Franzese, M. I. Marques, and H. E. Stanley, Phys. Rev. E {\bf
67}, 011103 (2003). 

\bibitem{fs2} G. Franzese and H. E. Stanley, 
J. Phys.: Cond. Mat., in print (2007).

\bibitem{slt}
F. Sciortino, E. La Nave, and P. Tartaglia,
Phys. Rev. Lett. {\bf 91}, 155701 (2003).

\bibitem{Brovchenko} I. Brovchenko, A. Geiger, and A. Oleinikova,
J. Chem. Phys. {\bf
123}, 044515 (2005).

\bibitem{MORISHITA}
T. Morishita, Phys. Rev. Lett. {\bf 87}, 105701 (2001).

\bibitem{Saika-Voivod}
I. Saika-Voivod, F. Sciortino and P. H. Poole, Phys. Rev. E 63,
011202-1 (2001).

\bibitem{SHRI_PP}
S. Sastry and C. A. Angell, Nature Mater. {\bf 2}, 739 (2003).

\bibitem{hydrogen}
S. Scandolo, Proc. Nat. Acad. Sci. USA {\bf 100} 3051 (2003).

\bibitem{Thiel}
M. van Thiel and F. H. Ree, Phys. Rev. B {\bf 48}, 3591 (1993).

\bibitem{G}
J. N. Glosli and F. H. Ree, Phys. Rev. Lett. {\bf 82}, 4659 (1999).

\bibitem{Ghiringhelli}
L. M. Ghiringhelli,  J. H. Los, E. J. Meijer,  A. Fasolino,
and D. Frenkel, Phys. Rev. Lett. {\bf 94}, 145701 (2005).

\bibitem{debenerev03}
P. Debenedetti,
J. Phys.: Cond. Mat. {\bf 15}, R1669 (2003).

\bibitem {hs}
P. C. Hemmer and G. Stell, Phys. Rev. Lett. {\bf 24}, 1284 (1970);
G. Stell and P. C. Hemmer, J. Chem. Phys. {\bf 56}, 4274 (1972).

\bibitem {Deb1}
P. G. Debenedetti, {\it Metastable Liquids: Concepts and
Principles} (Princeton University Press, Princeton, 1998); {\it
Hydration Processes in Biology. Theoretical and Experimental
Approaches}, Vol. 305 of NATO Advanced Studies Institute, Series
A: Life Sciences, edited by M. C. Bellissent-Funel (IOS
Press, Amsterdam, 1998).


\bibitem {SLK}
M. Silbert and W. H. Young, Phys. Lett. {\bf 58A}, 469 (1976);
D. Levesque and J. J. Weis, {\em ibid.} {\bf 60A}, 473 (1977); J. M.
Kincaid and G. Stell, {\em ibid}. {\bf 65A}, 131 (1978).

\bibitem {SA}
W. M. Shyu, J. H. Wehling, and M. R.Cordes, Phys. Rev. B {\bf 4},
1802 (1971); M. Appapillai and V.Heine, Cavendish Laboratory
Technical Report No. 5, 1972 (unpublished); K. K. Mon, N. W. Ashcroft,
and G. V. Chester, Phys. Rev. B {\bf 19}, 5103 (1979).

\bibitem{L75}
J. M. Lawrence, M. C. Croft, and R. D. Parks, Phys. Rev. Lett.
{\bf 35}, 289 (1975).

\bibitem{DRB91}
P. G. Debenedetti, V. S. Raghavan, and S. S. Borick, J. Phys.
Chem. {\bf 95}, 4540 (1991).

\bibitem{SHG93}
F. H. Stillinger and T. Head-Gordon, Phys. Rev. E {\bf 47}, 2484
(1993).

\bibitem{Deb2}
S. S. Borick, P. G. Debenedetti, and S. Sastry, J. Phys. Chem.
{\bf 99}, 3781 (1995); T. M. Truskett, P. G. Debenedetti, S.
Sastry, and S. Torquato, J. Chem. Phys. {\bf 111}, 2647 (1999).

\bibitem{ssbs}
M. R. Sadr-Lahijany, A. Scala, S. V. Buldyrev, and H. E. Stanley,
Phys. Rev. Lett. {\bf 81}, 4895 (1998); Phys. Rev. E {\bf 60},
6714 (1999); A. Scala, M. R. Sadr-Lahijany, N. Giovambattista,
S. V. Buldyrev, H. E. Stanley, {\em ibid.} {\bf 63}, 041202 (2001); J.
Stat. Phys. {\bf 100},97 (2000).

\bibitem{jagla}
E. A. Jagla, Phys. Rev. E {\bf 58}, 1478 (1998); 
J. Chem. Phys. {\bf 110}, 451 (1999);
J. Chem. Phys. {\bf 111}, 8980 (1999); 
Phys. Rev. E 63, 061501 (2001);
Phys. Rev. E {\bf 63}, 061509 (2001).

\bibitem{wilding}
N. B. Wilding and J. E. Magee,
Phys. Rev. E {\bf 66}, 031509 (2002);
H. M. Gibson, N. B. Wilding,
Phys. Rev. E {\bf 74}, 019903 (2006).

\bibitem{Caballero}
J. B. Caballero, A. M. Puertas.
Phys. Rev. E {\bf 74}, 051506 (2006).

\bibitem{StillingerJCP78}
F. H. Stillinger and T. A. Weber,
J. Chem. Phys. {\bf 68} 3837
1978.

\bibitem{stanley}
P. Kumar, S. V. Buldyrev, F. Sciortino, E. Zaccarelli, and H. E. Stanley,
Phys. Rev. E {\bf 72}, 021501 (2005);
Z. Yan, S. V. Buldyrev, N. Giovambattista, and H. E. Stanley,
Phys. Rev. Lett. {\bf 95}, 130604 (2005);
L. Xu, P. Kumar, S. V. Buldyrev, S.-H. Chen, P. Poole, F. Sciortino,
and H. E.  Stanley,
Proc. Natl. Acad. Sci. U.S.A. {\bf 102}, 16558 (2005);
L. Xu, S. Buldyrev, C. A. Angell, and H. E. Stanley,
Phys. Rev. E {\bf 74}, 031108 (2006);
Z. Yan, S. V. Buldyrev, N. Giovambattista, P. G. Debenedetti, and H. E.
  Stanley,
Phys. Rev. E {\bf 73}, 051204 (2006).

\bibitem{sharma}
R. Sharma, A. Mudi, and C. Chakravarty,
J. Chem. Phys. {\bf 125}, 044705 (2006).
R. Sharma, S. N. Chakraborty, and C. Chakravarty,
J. Chem. Phys. {\bf 125}, 204501 (2006).

\bibitem{camp}
P. Camp,
Phys. Rev. E {\bf 68}, 061506 (2003);
Phys. Rev. E {\bf 71}, 031507 (2005).

\bibitem{esposito}
R. Esposito, F. Saija, A. M. Saitta, P. V. Giaquinta,
Phys. Rev. E {\bf 73}, 040502 (2006).

\bibitem{deoliveira}
A. B. de Oliveira, P. A. Netz, T. Colla, and M. C. Barbosa,
J. Chem. Phys. {\bf 124}, 084505 (2006);
J. Chem. Phys. {\bf 125}, 124503 (2006).

\bibitem{mittal}
J. Mittal, J. R. Errington, and T. M. Truskett,
J. Chem. Phys. {\bf 125}, 076102 (2006).

\bibitem{nature} G. Franzese, G. Malescio, A. Skibinsky, S.V. Bulderev,
and H. E. Stanley, Nature {\bf 409}, 692 (2001).

\bibitem{fmsbs} G. Franzese, G. Malescio, A. Skibinsky, S. V. Buldyrev,
and H. E. Stanley, Phys. Rev. E {\bf 66}, 051206 (2002).

\bibitem{sbfms}
A. Skibinsky, S.V. Buldyrev, G. Franzese, G. Malescio, and H. E. Stanley,  
Phys. Rev. E {\bf 69}, 061206 (2004).

\bibitem{mfsbs}
G. Malescio, G. Franzese, A. Skibinsky, S. V. Buldyrev, and H. E. Stanley,  
Phys. Rev. E {\bf 71}, 061504 (2005).

\bibitem{barbosa}
A. L. Balladares and M. C. Barbosa,
J. Phys.: Cond. Matter {\bf 16}, 8811 (2004).
A. B. de Oliveira and M. C. Barbosa,
J. Phys.: Cond. Matter {\bf 17}, 399 (2005).
V. B. Henriques and M. C. Barbosa,
Phys. Rev. E {\bf 71}, 031504 (2005).
V. B. Henriques, N. Guissoni, M. A. Barbosa, M. Thielo, and M. C. Barbosa,
Mol. Phys. {\bf 103}, 3001 (2005).

\bibitem{buldyrev}
A. Scala, M. R. Sadr-Lahijany, N. Giovambattista, S. V. Buldyrev, and
H. E. Stanley,
J. Stat. Phys. {\bf 100}, 97 (2000);
Phys. Rev. E {\bf 63}, 041202 (2001).
S. V. Buldyrev, G. Franzese, N. Giovambattista, G. Malescio, M. R.
Sadr-Lahijany, A. Scala, A. Skibinsky, and H. E. Stanley,
Physica A {\bf 304}, 23 (2002).
S. V. Buldyrev and H. E. Stanley,
Physica A {\bf 330}, 124 (2003).

\bibitem{WaxPRB00}
See for example,
F. Wax, R. Albaki, and J.-L. Bretonnet,
Phys. Rev. B {\bf 62}, 14818 
(2000).

\bibitem{DebenedettiJPC91}
P. G. Debenedetti, V. S. Raghavan, and S.S. Borick,
J. Phys. Chem. {\bf 95}, 4540 
(1991) 
and references therein.

\bibitem{QuesadaJCP01}
M. Quesada-P\'erez, A. Moncho-Jord\'a, F. Mart\'{\i}nez-L\'opez, and
R. Hidalgo-\'Alvarez
J. Chem. Phys. {\bf 115}, 10897 
(2001).

\bibitem{ChoPRL}
C. H. Cho, Surjit Singh, and G. W. Robinson
Phys. Rev. Lett. 76, 1651 
(1996).

\bibitem{NetzPhysA04}
P. A. Netz, J. F. Raymundi, A. S. Camera, and M. C. Barbosa,
Physica A {\bf 342}, 48 
(2004).

\bibitem{AT}
M.P. Allen and D.J. Tildesley, {\it Computer Simulation of
Liquids}, Oxford University Press, New York, 1989.

\bibitem{debene-stanley-review}
P. G. Debenedetti and H. E. Stanley, Physics Today {\bf 56} 
40 (2003).

\end{thebibliography}
\end{document}